\documentclass[conference]{IEEEtran}
\IEEEoverridecommandlockouts
\usepackage{cite}
\usepackage{url}
\usepackage{amsmath,amssymb,amsfonts}
\usepackage{algorithmic}
\usepackage{graphicx}
\usepackage{textcomp}
\usepackage{xcolor}
\usepackage{multicol}
\usepackage{stfloats}
\usepackage{multirow}
\def\BibTeX{{\rm B\kern-.05em{\sc i\kern-.025em b}\kern-.08em
    T\kern-.1667em\lower.7ex\hbox{E}\kern-.125emX}}

\begin{document}

\title{Smart Contract Generation for Inter-Organizational Process Collaboration\\
\thanks{.}
}

\author{\IEEEauthorblockN{1\textsuperscript{st} Tianhong Xiong\textsuperscript{1,2}}
\IEEEauthorblockA{\textit{1.School of Computer Science and Engineering} \\
\textit{Sun Yat-sen University}\\
Guangzhou, China \\
\textit{2.School of Digital Media and Humanities} \\
\textit{Hunan University of Technology and Business}\\
Changsha, China \\
xiongth3@mail2.sysu.edu.cn}
\and
\IEEEauthorblockN{2\textsuperscript{nd} Shangqing Feng}
\IEEEauthorblockA{\textit{School of Computer Science and Engineering} \\
\textit{Sun Yat-sen University}\\
Guangzhou, China \\
fengshq5@mail2.sysu.edu.cn}
\and
\IEEEauthorblockN{3\textsuperscript{rd} Maolin Pan}
\IEEEauthorblockA{\textit{School of Computer Science and Engineering} \\
\textit{Sun Yat-sen University}\\
Guangzhou, China \\
panml@mail.sysu.edu.cn}
\and
\IEEEauthorblockN{4\textsuperscript{th} Yang Yu*}
\IEEEauthorblockA{\textit{School of Computer Science and Engineering} \\
\textit{Sun Yat-sen University}\\
Guangzhou, China \\
yuy@mail.sysu.edu.cn}
}
\maketitle

\begin{abstract}
Currently, inter-organizational process collaboration (IOPC) has been widely used in the design and development of distributed systems that support business process execution. Blockchain-based IOPC can establish trusted data sharing among participants, attracting more and more attention. The core of such study is to translate the graphical model (e.g., BPMN) into program code called smart contract that can be executed in the blockchain environment. In this context, a proper smart contract plays a vital role in the correct implementation of block-chain-based IOPC. In fact, the quality of graphical model affects the smart con-tract generation. Problematic models (e.g., deadlock) will result in incorrect contracts (causing unexpected behaviours). To avoid this undesired implementation, this paper explores to generate smart contracts by using the verified formal model as input instead of graphical model. Specifically, we introduce a prototype framework that supports the automatic generation of smart contracts, providing an end-to-end solution from modeling, verification, translation to implementation. One of the cores of this framework is to provide a CSP\#-based formalization for the BPMN collaboration model from the perspective of message interaction. This formalization provides precise execution semantics and model verification for graphical models, and a verified formal model for smart contract generation. Another novelty is that it introduces a syntax tree-based translation algorithm to directly map the formal model into a smart contract. The required formalism, verification and translation techniques are transparent to users without imposing additional burdens. Finally, a set of experiments shows the effectiveness of the framework.
\end{abstract}


\section{Introduction}
At present, inter-organizational process collaboration (IOPC), as a distributed para-digm, has been widely used in the design and development of software systems that support business process execution \cite{ref1}. Furthermore, because of its powerful modeling ability, Business Process Model and Notation 2.0 (BPMN for short) \cite{ref2} has become one of the most commonly used proposals when defining process models, supporting the design and implementation of IOPC from the perspective of model-driven development.

Blockchain is an emerging decentralized technology for establishing trusted data sharing among participants \cite{ref3}. Since it is tamper-proof, the executed log records will not be disputed by possible forgery by participants or third parties. This enables it to provide a full process audit trail of transactional data. Benefiting from the above features, IOPC combined with blockchain has attracted more and more attention, and has been explored in many fields, e.g., supply chain [4, 5], government service \cite{ref6}, e-commerce \cite{ref7} and others [8-10]. This type of study usually translates the model (e.g., BPMN) into program code called smart contract that can be executed in the blockchain environment, and coordinates and records the task execution and interaction of the participants. In this context, a smart contract that meets expectations plays a vital role in the correct implementation of blockchain-based IOPC.

In fact, the generation of smart contract (as object code) is directly affected by the graphical model of IOPC (as source code). Although BPMN has been widely used as a de facto modeling standard, semiformal definition and natural text description sometimes contain misleading information in model description \cite{ref11}. This may become more serious in industrial environment, because model designers are usually not familiar with formalism and verification technology, but are used to standard graphical symbols. In this context, the quality of the graphical models may be uneven. The low-quality models (e.g., deadlock) may result in incorrect contracts that cause undesirable behaviours, and more importantly, it is hard to repair the contracts after implementation. 

To avoid undesired implementation, this paper proposes an automated framework support the smart contract generation for IOPC. It covers the complete life cycle of IOPC from modeling, verification, translation to implementation. One of its core is to propose a suitable formal model as the source code for smart contract generation. As we all know, a formal model has clear execution semantics, can accurately describe the behavior specification, and avoid the ambiguity caused by semiformal and text description. In particular, it can identify problematic models in advance through model checking, which enhances the confidence of quality of models and software systems. These are particularly important for IOPC involving a large number of message interactions, because the graphical model is prone to deadlock due to message congestion, and its corresponding smart contract is difficult to repair in real time \cite{ref8}. 

Specifically, the framework first uses BPMN collaboration model to represent IOPC. In practice, BPMN collaboration can intuitively representing the boundaries and business responsibilities of participants, focusing on interaction in collaboration [3, p. 317]. Furthermore, due to the advantages of Communicating Sequential Programs (CSP\#) \cite{ref12} in message communication and structured representation, we give a structured formalization for the BPMN model and the corresponding automatic translation algorithm, that is, the model elements are mapped to the corresponding composition of CSP\# processes. This facilitates the subsequent smart contract generation.

Another key is the translation from CSP\# model to smart contract. The formal model is parsed into a syntax tree structure, in which all internal nodes are non-terminal symbols, and each leaf node is a terminal symbol, representing a component element of the CSP\# model. Based on the syntax tree, a two-stage translation algorithm is given, with a formal model as input instead of a semiformal graphical model. It first traverses the relationships among the formal model elements to determine the logical execution order, and then translates each element in turn into the smart contract codes written in Solidity. Finally, a set of experiments shows the effectiveness of prototype framework.

In summary, the main contributions of this paper include the followings:

\begin{itemize}
    \item We propose a prototype framework for automatic smart contract generation for IOPC, which provides an end-to-end solution integrating "modeling-verification-translation-implementation". The required formalism and translation technology are transparent to users, so the framework does not impose additional user burden while achieving the expected goal.
    \item We give a CSP\# formalization of BPMN collaboration model and the corresponding translation algorithm from BPMN model to formal CSP\# model. Different from the existing proposals, it is structured and focuses on message communication, ignoring internal elements that do not participate in interaction. This structured formalization is particularly suitable for mapping to smart contract, reducing the complexity of translation.
    \item We develop a translation algorithm from CSP\# model to smart contract. It starts with defining the association relationships between model elements, and then generates contract code by parsing the syntax tree of formal model. This algorithm employs the knowledge of compilation principles (lexical and syntax analysis), which is conducive to accurate understanding and translation of smart contract.
\end{itemize}

The rest of this paper is organized as follows. Section 2 briefly describes a running example to illustrate the context and motivation of this work. Section 3 introduces the framework and Section 4 describes the core methods in detail. Section 5 shows a set of experiments and discusses some observations obtained in the experiments. After reviewing the related work in Section 6, Section 7 summarizes the paper.

\section{Running Example}
This section introduces a supply chain process collaboration scenario as an example throughout this paper, which is convenient to explain the context and motivation of our work. 

In the collaboration scenario, the participants come from different organizations and assume different business responsibilities. Each participant has its own BPMS and establishes interaction with each other through the blockchain-based collaboration service (agent). The internal processes of participants are implemented on its own BPMS to ensure privacy, while the collaboration process is translated into a Solidity smart contract deployed in the blockchain environment, which stores transaction information and controls the interaction order to provide mutual trust.

\begin{figure*}[htbp]
    \centerline{\includegraphics[width=0.97\linewidth]{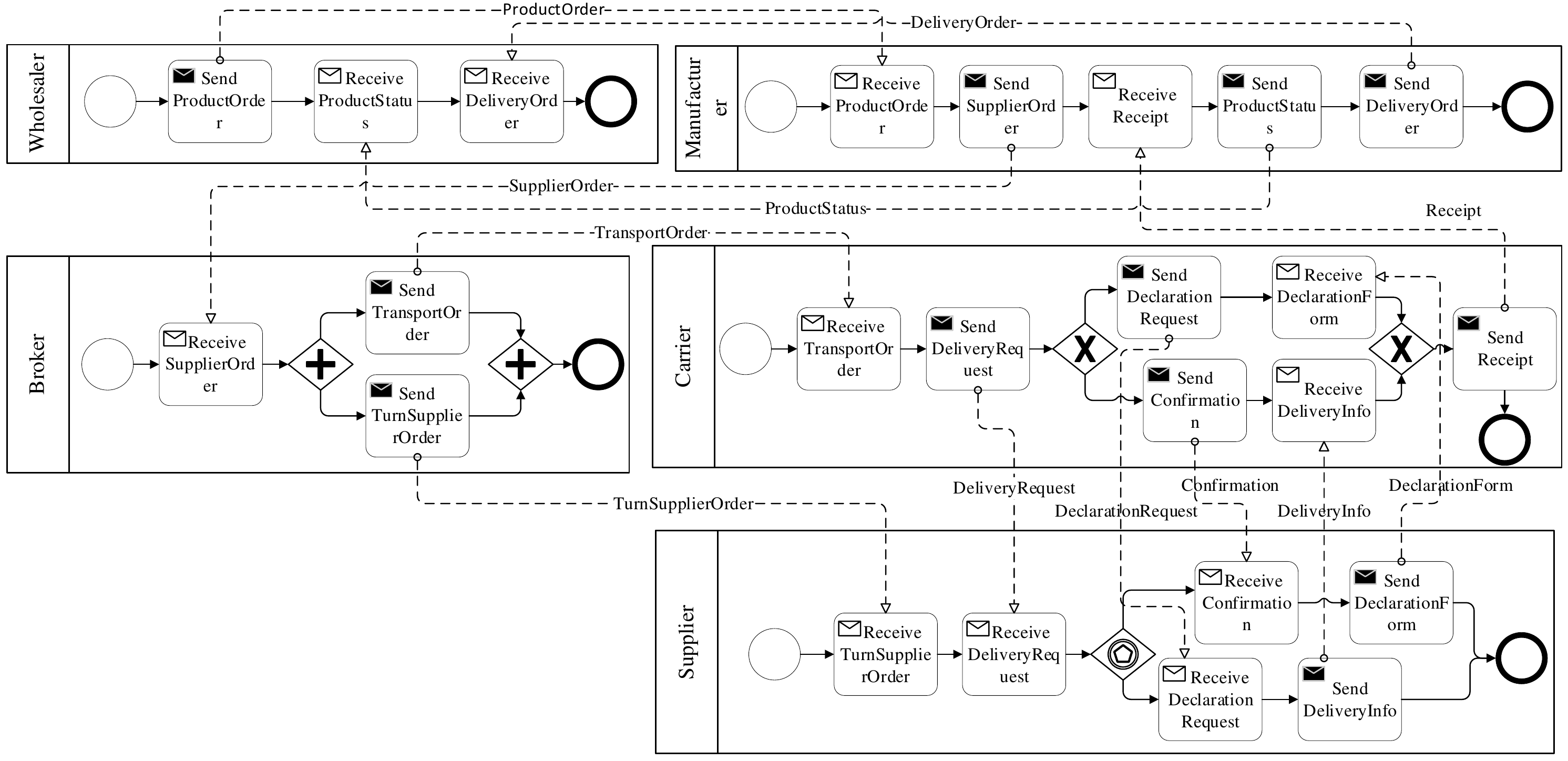}}
    \caption{BPMN Collaboration model of supply chain scenario.}
    \label{fig:figure1}
\end{figure*}

The supply chain example adapted from literature \cite{ref4} describes the manufacturing and delivery process of product order, involving 5 participants: \textit{Wholesaler}, \textit{Manufacturer}, \textit{Broker}, \textit{Supplier} and \textit{Carrier}. The collaboration model is shown in Fig.~\ref{fig:figure1}. We don't provide a detailed description of the model here, because the meaning of each task should be intuitive. 

This collaboration scenario needs to fully consider the model design of multiple participants and a large number of message interactions among participants. Indeed, complex model design and message flow configuration is an error prone task. For example, in the collaboration between \textit{Carrier} and \textit{Supplier}, messages are blocked due to the wrong configuration of tasks and messages in the gateways, resulting in deadlock. Once this model is translated and deployed directly, blockchain may lose its essential meaning, that is, it cannot achieve accurate collaboration with this supply chain scenario. More detailed knowledge about BPMN, CSP\# and blockchain is provided in the technical report of the repository\footnote{https://github.com/xthHub/SCG4IOPC}.

\section{Framework}
\begin{figure*}[htbp]
    \centerline{\includegraphics[width=0.8\linewidth]{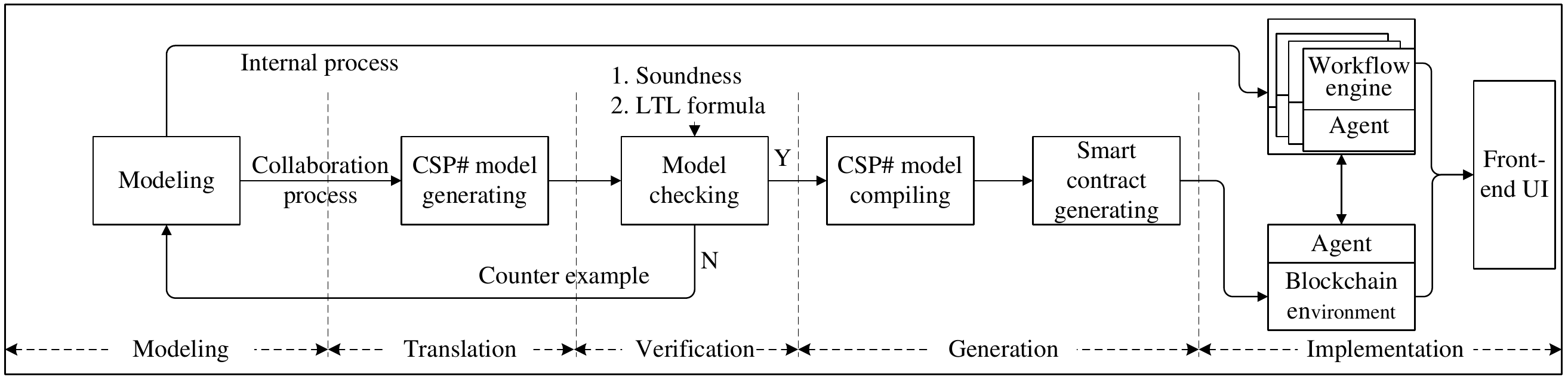}}
    \caption{The overview of integrated framework.}
    \label{fig:figure2}
\end{figure*}

The framework provides an end-to-end solution from modeling to implementation for smart contract generation for IPOC. It consists of 5 parts: Modeling, Translation (translate BPMN to CSP\# model), Verification, Generation (translate CSP\# model to smart contract), and Implementation. Fig.~\ref{fig:figure2} shows the internal components of the framework and the user interfaces/operations related to the system designers. 

The special feature of framework is that the system designers only need to focus on the modeling, optimization (e.g., adjusting the model according to the counter examples) and monitoring, without mastering the formal language, verification and parsing techniques. The framework is developed as a stand-alone solution, but it can also be integrated as a service accessed through RESTful interface or as a plug-in in existing tools. 

The modeling environment integrates a graphical modeling tool Camunda\_bpmn.io \cite{ref13} that supports BPMN standard. Furthermore, considering the diversity of existing modeling tools, the framework does not impose any restrictions on the source of the graphical models. Input models (.bpmn file format) can be created by system designers using different BPMN modeling tools, or retrieved from a common repository. In this regard, the framework is not limited to specific modeling tools, and is easy to use by system designers from different business fields. 

The framework provides an implementation environment, as shown in the right part of Fig.~\ref{fig:figure2}. Workflow engines (BPMS) are logically aggregated and physically dispersed. They perform the internal processes of participants and interact with the Blockchain environment through Agents. The Solidity smart contract deployed in blockchain is used to record transaction data and control the order of interaction among participants. Here, workflow engine extends the open source engine Zeebe\footnote{https://github.com/camunda-cloud/zeebe} to support the instantiation of BPMN model. Blockchain environment is based on Ganache\footnote{https://www.trufflesuite.com/ganache} that is a node emulator of Ethereum. The agent based on the interoperability protocol Wf-XML 2.0 \cite{ref14}, is responsible for monitoring the state changes, initiating or receiving (actively subscribing) external requests, and triggering task or event execution. The Front-end UI provides a user-friendly web interface and supports the complete data display.

\section{Methods}
This section introduces CSP\# model translation and smart contract generation in detail.
\subsection{Translation from BPMN to CSP\# Model and Formal Verification}
This sub section describes the translation of BPMN to CSP\# model, and the property verification. To simplify CSP\# formalization, inspired by literature \cite{ref11}, we apply the general syntax notation BNF to textual BPMN model.

\begin{figure*}[htbp]
    \centerline{\includegraphics[width=0.6\linewidth]{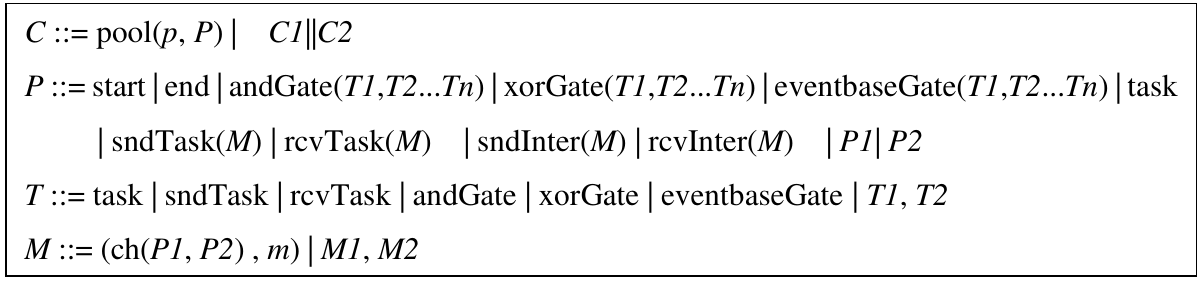}}
    \caption{The overview of integrated framework.}
    \label{fig:figure3}
\end{figure*}

\paragraph{Translation}The BNF syntax definition of the core elements of the collaboration model is given, as shown in Fig.~\ref{fig:figure3}. It maps model elements to structured text descriptions to facilitate subsequent formalization. In this syntax, \textit{C}, \textit{P}, \textit{T} and \textit{M} represent collaboration structure, participant structure, element list and message list respectively.

A collaboration \textit{C} represents a model composition that associates \textit{pool (p, P)} with the parallel operator "$||$". In the pool, \textit{p} is the name of the pool and \textit{P} is the encapsulated participant model (process). To simplify formalization, we assume that each pool contains only one participant, and treat the gateway as a structured whole containing split and join mode, with unique input and output sequence flow. For example, andGate (\textit{T1}, \textit{T2}...\textit{Tn}) captures all the elements it affects, where \textit{T1}, \textit{T2}...\textit{Tn} respectively represent the element list on each inner parallel edge. The message list \textit{M} contains a series of triple (\textit{ch(P1, P2) , m}), where \textit{ch} is the channel, \textit{P1} is the sender, \textit{P2} is the receiver, and m is the designated unique message name.

The above BNF syntax only provides a structured description. To describe the semantics, we employ the features of structured representation supported by CSP\#. Each graphical model element is mapped to a structured CSP\# process consisting of alphabet, operators and keywords, as shown in Table 1. In task element, $event_{work}$ represents the work of the task (i.e., \textit{$event_{work}$ -$>$ Skip};). We assume that the send and receive tasks (i.e., \textit{sndTask} and \textit{rcvTask}) are only responsible for message interaction. According to task types, $event_{work}$ can be mapped to channel operations (i.e., \textit{ch}!\textit{m} and \textit{ch}?\textit{m}, respectively). Note that the channel can be seen as a FIFO queue here.

\begin{table*}[]
    \centering
    \caption{A CSP\# formalization for core collaboration model elements}
    \begin{tabular}{|c|l|l|}
    \hline
    Elements                  & \multicolumn{1}{c|}{BPMN Syntax}    & \multicolumn{1}{c|}{CSP\#}                                                                    \\ \hline
    \multirow{3}{*}{Tasks}    & task($e_{i}$, $e_{o}$)                        & $event_{ei}$ -\textgreater Skip; $event_{work}$ -\textgreater Skip; $event_{eo}$ -\textgreater Skip          \\ \cline{2-3} 
                              & sndTask($e_{i}$, M, $e_{o}$)                  & $event_{ei}$ -\textgreater Skip; ch!m -\textgreater Skip; $event_{eo}$ -\textgreater Skip               \\ \cline{2-3} 
                              & rcvTask($e_{i}$, M, $e_{o}$)                  & $event_{ei}$ -\textgreater Skip; ch?m -\textgreater Skip; $event_{eo}$ -\textgreater Skip               \\ \hline
    \multirow{3}{*}{Gateways} & andGate($e_{i}$, (T1,T2...Tn), $e_{o}$)       & $event_{ei}$ -\textgreater Skip; (T1 $||$ T2 $||$ ... $||$  Tn); $event_{eo}$ -\textgreater Skip                  \\ \cline{2-3} 
                              & xorGate($e_{i}$, (T1,T2...Tn), $e_{o}$)       & $event_{ei}$ -\textgreater Skip; (T1 {[}{]} T2 {[}{]}… {[}{]} Tn) ); $event_{eo}$ -\textgreater Skip    \\ \cline{2-3} 
                              & eventbaseGate($e_{i}$, (T1,T2...Tn), $e_{o}$) & $event_{ei}$ -\textgreater Skip; (T1 {[}*{]} T2 {[}*{]}… {[}*{]} Tn) ); $event_{eo}$ -\textgreater Skip \\ \hline
    \end{tabular}
    \label{tab:table1}
    \end{table*}

The gateway affects multiple elements. These elements (e.g., multiple tasks) are associated by the sequential, parallel or choice operators. Thus, according to the covered elements, the gateway is represented as a composition of elements. For example, the CSP\# process of \textit{Broker} participant (see Fig.~\ref{fig:figure1}) is a composite process that contains three task elements. 


\textit{Broker}() =(\textit{$event_{e1}$} -$>$ \textit{Skip}; \textit{cMB}?\textit{SupplierOrder} -$>$ \textit{Skip}; \textit{$event_{e2}$} -$>$ \textit{Skip}); ((\textit{$event_{e2}$} -$>$ \textit{Skip}; \textit{cBS}!\textit{TurnSupplierOrder}-$>$ \textit{Skip}; \textit{$event_{e3}$} -$>$ \textit{Skip}) $||$ (\textit{$event_{e2}$} -$>$ \textit{Skip}; \textit{cBC}!\textit{TransportOrder} -$>$ \textit{Skip}; \textit{$event_{e3}$} -$>$ \textit{Skip})). 

This means that the \textit{Broker} first receives message \textit{SupplierOrder} through channel \textit{cMB}, and then it will perform the parallel operation, sending messages \textit{TurnSupplierOrder} and \textit{TransportOrder} through channels \textit{cBS} and \textit{cBC}, respectively.

The corresponding translation algorithm first reads the BPMN model and stores the elements in the corresponding nodes. Then it traverses the model and constructs NextElements. Finally, it adopts depth first traversal method to recursively maps elements to CSP\# process according to the rules in Table 1. Limited by space, only part of the algorithm is shown below.

\begin{figure}
    \centering
    \includegraphics[width=1\linewidth]{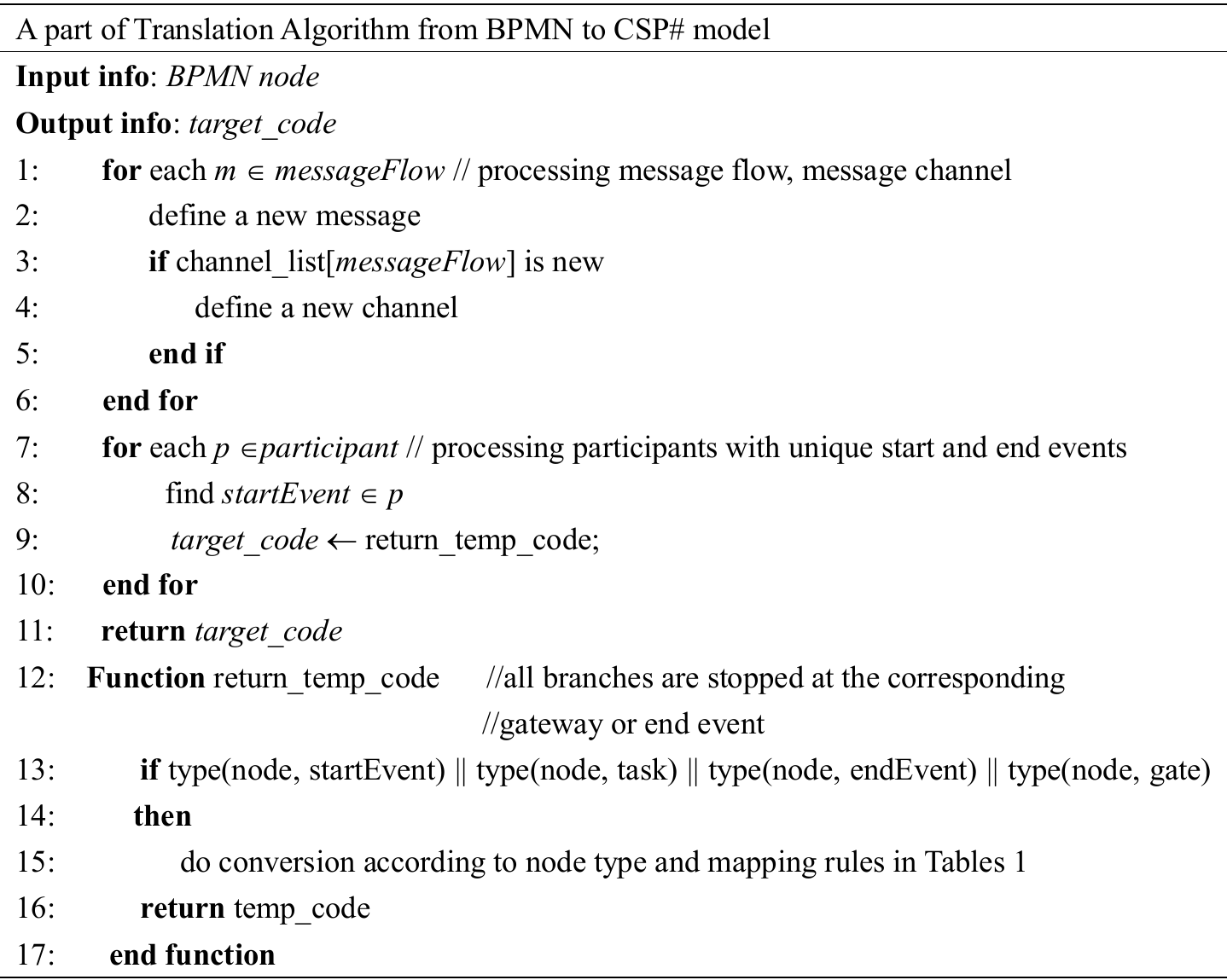}
    \label{alg:chapter3-2}
  \end{figure}

\paragraph{Formal Verification}This paper focuses on the soundness verification of collaboration model. It is a fundamental requirement for IOPC \cite{ref15}. The IOPC is sound if all participants are sound and there are no undelivered messages among participants. The soundness of participants means that it is terminable, there is no deadlock, and each task is reachable (executable). The formal definition is given below.

Formally, Soundness: $\forall$ \textit{participant} $\in$ \textit{P}: deadlockfree(\textit{participant}) $\wedge$ $<>$ ($\forall$ \textit{participant} $\in$ \textit{P}: \textit{participant} reaches \textit{end} $\wedge$ $\forall$ \textit{ch} $\in$ channel: $|ch|$=0), where \textit{deadlockfree}, \textit{nonterminating} and \textit{reaches} are CSP\# based attribute assertions, \textit{end} is a given conditional proposition and “$|ch|$ =0” indicates that there is no message in the channel \textit{ch}.

In fact, besides soundness, user-defined property verification is also supported. For example, given a participant, the following assertion asks whether participant satisfies the LTL formula. 
Formally, \#assert \textit{participant} $|$= \textit{F}, where \textit{F} is the LTL formula.

According to the selected properties, the framework automatically completes formal verification. Once the verification fails, it will feedback a counter example.

\subsection{Translation from CSP\# Model to Smart Contract}
This section introduces the translation algorithm of mapping CSP\# model to Solidity smart contract. Inspired by the knowledge of lexical and syntax analysis, a two-stage algorithm is given. Firstly, a set of association relationships is defined to specify the logical order among the elements within the participants and among the participants. Secondly, the syntax tree of the formal model is given, and a traversal algorithm is used to capture the set of association relationships based on the syntax tree.

\paragraph{Association relationships between CSP\# Processes}Association relationships describe the interaction (execution order) among different model elements within participants and among participants. These relationships are the basis of automatic generation of smart contracts. Here we defines 6 kinds of relationships, including Next, End, Init, And, Xor and Enable. The first five of them indicates the interaction among the CSP\# processes within the participants, while the last one specifies the interaction logic constraints among the participants. The 6 relationships are essentially the analysis of process structure. They cover sequence, branch, parallel and loop, and can combine any single entry and single exit process structure.

\begin{itemize}
    \item Next describes the sequential relationship between CSP\# processes within a participant. If CSP\# process \textit{P1} ends, then CSP\# process \textit{P2}[] will start. That is, Next (\textit{P1}) = [\textit{P2}[]] where \textit{P2}[] is a CSP\# process array.
    \item End describes the end relationship between CSP\# processes within a participant. When a process ends, it triggers other processes to end. If \textit{P1} ends, then \textit{P2}[] will end. Namely, End (\textit{P1}) = [\textit{P2}[]].
    \item Init describes the initialization relationship between CSP\# processes within a participant. If \textit{P1} begin, then \textit{P2}[] begins first. That is, Init (\textit{P1}) = [\textit{P2}[]].
    \item And describes the relationship among CSP\# processes involved in parallel gateway. If And (\textit{P1}) = [\textit{P2}[]], then \textit{P1} ends only if \textit{P2}[] ends.
    \item Xor describes the relationship among CSP\# processes involved in exclusive gateway. If \textit{P2} $\in$ Xor (\textit{P1}) is executing, $\forall$ \textit{P3} $\in$ Xor (\textit{P1}) is disabled.
    \item Enable describes the enabling relationship of message interaction between participants. The message flow \textit{M} specifies the enabling relationship between sender and receiver. If \textit{M} = (ch(\textit{P1}, \textit{P2}), m), Enable (\textit{P1}) = [\textit{P2}].
\end{itemize}

\begin{figure}[htbp]
    \centerline{\includegraphics[width=0.95\linewidth]{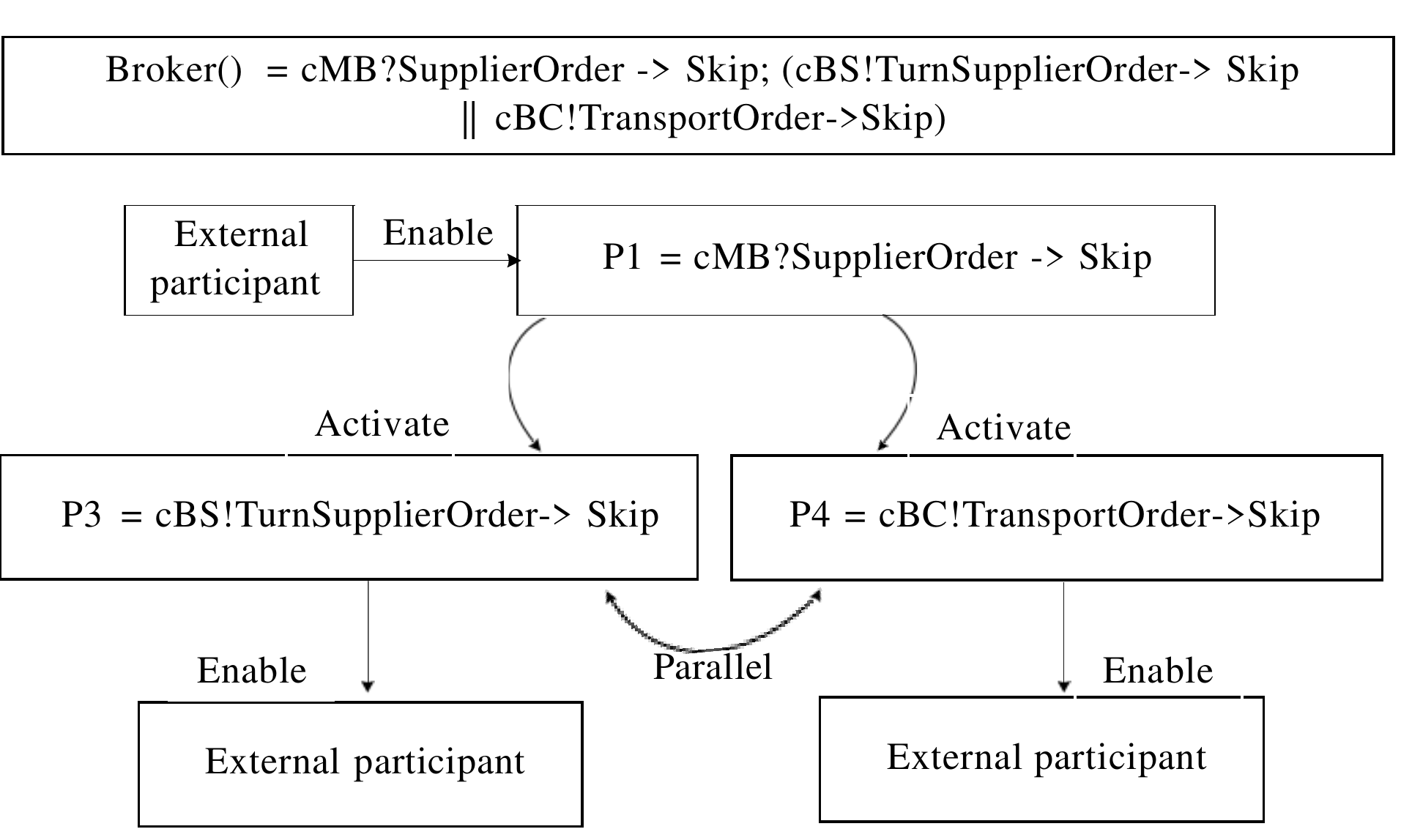}}
    \caption{The relationships of \textit{Broker} participant.}
    \label{fig:figure4}
\end{figure}

Fig.~\ref{fig:figure4} shows the relationships of \textit{Broker} participant in the supply chain example. \textit{Broker} is a \textbf{composite process}, so is \textit{P2}, which is composed of more than one \textbf{atomic process} like \textit{P1}, \textit{P3} and \textit{P4}. 

Init (\textit{Broker}) = [\textit{P1}], which means that when the \textit{Broker} is started, \textit{P1} will execute first. Next (\textit{P1}) = [\textit{P2}], which means that when \textit{P1} is completed, \textit{P2} can start. End (\textit{P2}) = [\textit{Broker}], which means that when \textit{P2} is completed, it triggers the \textit{Broker} to end. In addition to complying with the internal relationships in \textit{Broker}, whether \textit{P1} (responsible for receiving \textit{SupplierOrder} message) can be executed also depends on the Enable relationship of the external participant (i.e., Enable (external participant) = [\textit{P1}]), because it requires the external participant to complete the sending of the message first. Because the focus of this paper is process collaboration, in order to facilitate a clear analysis of the collaboration model, here we ignore the internal events that do not participate in the interaction.

\begin{figure}[htbp]
    \centerline{\includegraphics[width=0.85\linewidth]{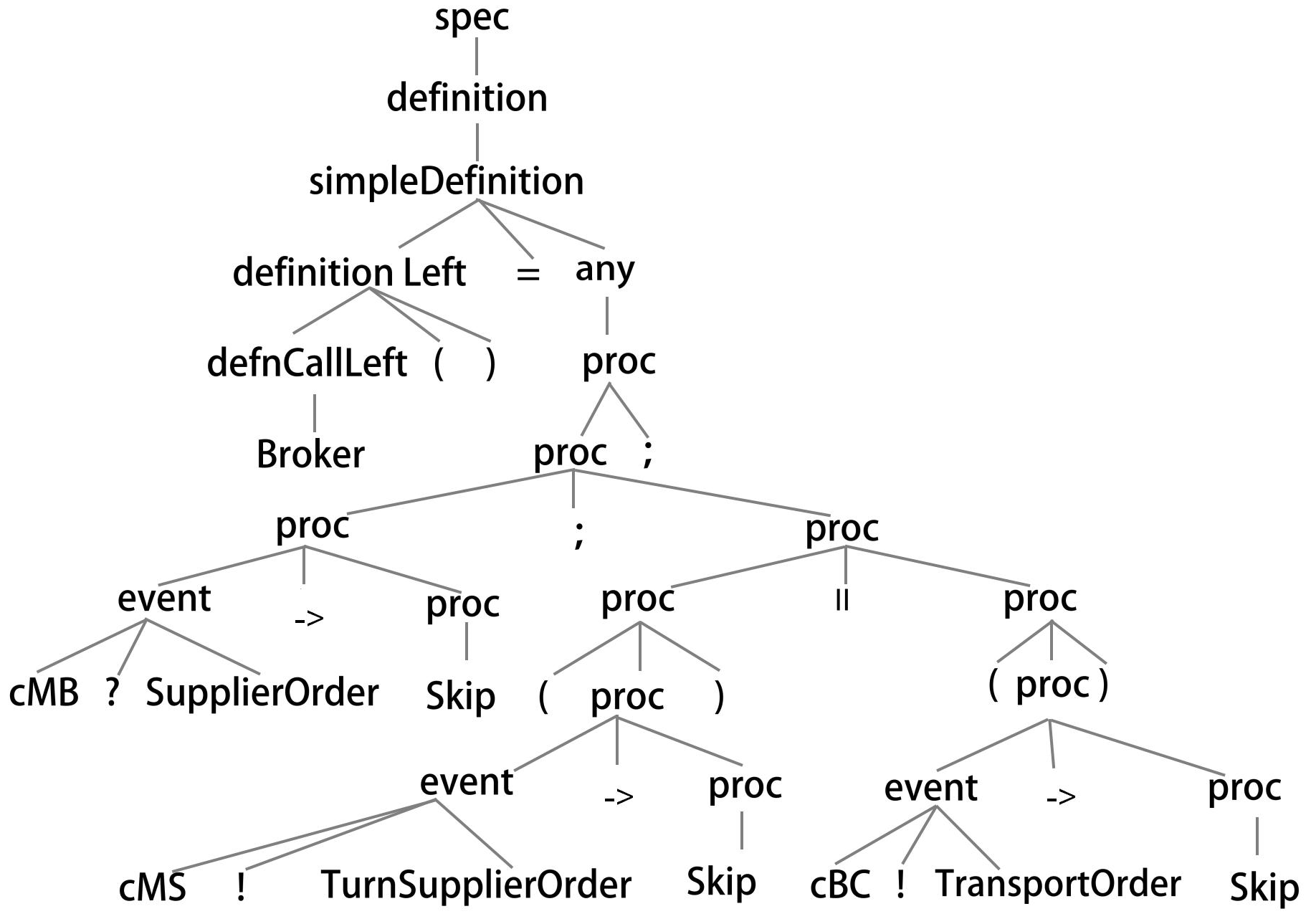}}
    \caption{The syntax tree of \textit{Broker} participant.}
    \label{fig:figure5}
\end{figure} 

\paragraph{Relationship traversal algorithm}

The relationship traversal algorithm is based on syntax tree. To facilitate the traversal and display of relationships, an syntax analysis tool ANTLR (Another Tool for Language Recognition) \cite{ref16} is used to obtain the syntax tree of CSP\# model. For example, Fig.~\ref{fig:figure5} shows the syntax tree of the \textit{Broker} participant, in which the leaf nodes form its CSP\# process from left to right, i.e., \textit{Broker}() = \textit{cMB}?\textit{SupplierOrder} -$>$ \textit{Skip}; (\textit{cBS}!\textit{TurnSupplierOrder}-$>$ \textit{Skip} $||$ \textit{cBC}!\textit{TransportOrder} -$>$ \textit{Skip}). The spec, definition, simpleDefinition, definitionLeft, defnCallLeft are the default reserved words. A part of the algorithm is given here as an example. It traverses the syntax tree to get the logical association relationships in the CSP\# processes.

\begin{figure}[htbp]
    \centerline{\includegraphics[width=1\linewidth]{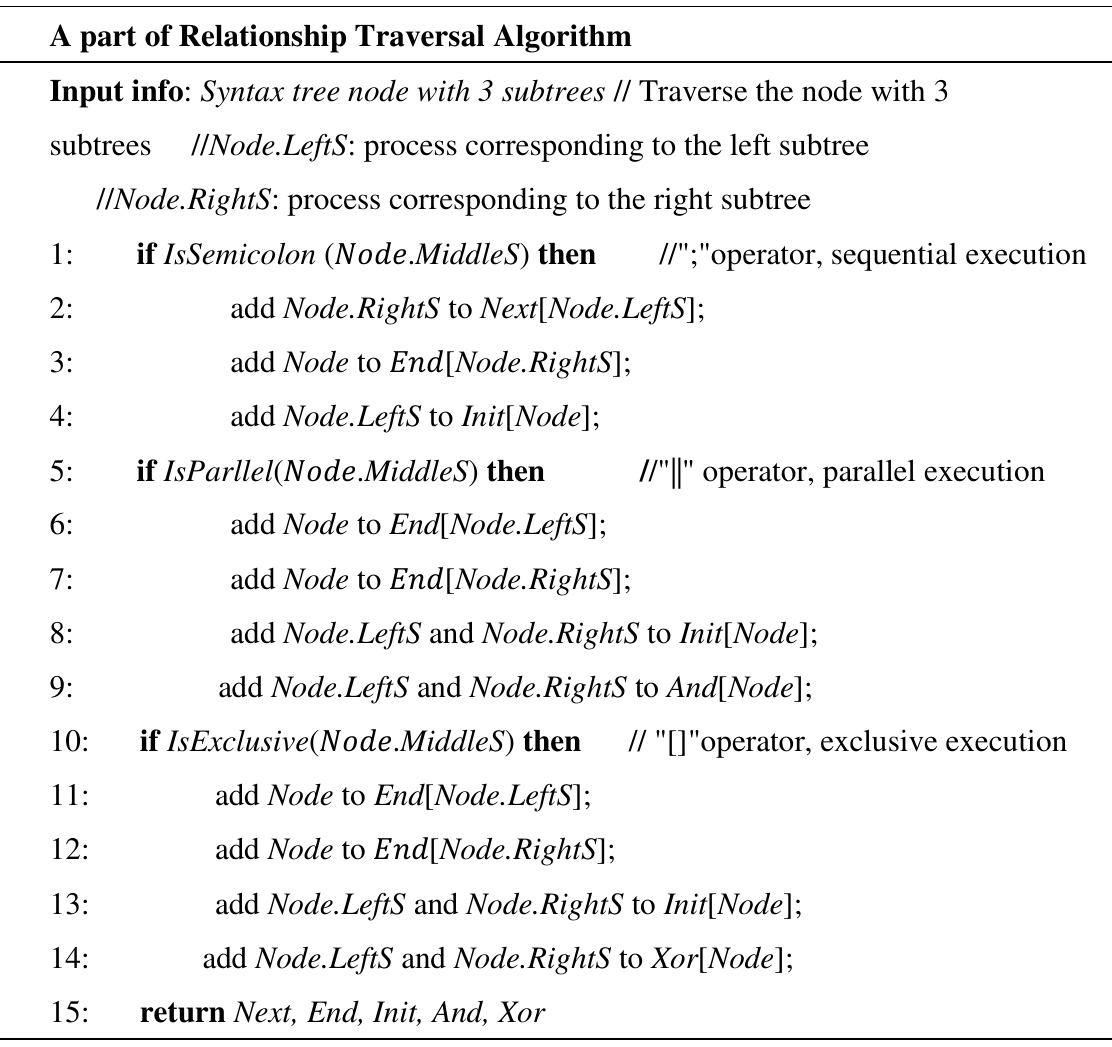}}
    \label{fig:figure6}
\end{figure} 

\paragraph{Reduction}

Our goal is to generate smart contracts, and the content of the contracts will affect the cost to execute them, so we take some measures to reduce the cost as much as possible. Note that the relationships defined above are used to describe the association of \textbf{all nodes} in the syntax tree, which contains many non-leaf nodes, corresponding to the CSP\# composite processes. However, leaf nodes are enough to form the whole CSP CSP\# composite process like \textit{Broker}(). So our strategy is to concentrate on the CSP\# \textbf{atomic processes} (leaf nodes in the syntax trees), and the relationships between them. Here we define three new relationships to achieve our reduction: Activate, Inactivate, Parallel. 

\begin{itemize}
    \item Activate describes the sequential relationship between CSP\# atomic processes within a participant. If atomic process \textit{P1} finishes its execution, atomic processes in \textit{P2}[] will start, then we have Activate (\textit{P1}) = [\textit{P2}[]] where \textit{P2}[] is a atomic process array.
    \item Inactivate describes the inactivate relationship between CSP\# atomic processes within a participant. If atomic process \textit{P1} finishes its execution, atomic processes in \textit{P2}[] can't be executed in this business process instance, then we have Inactivate (\textit{P1}) = [\textit{P2}[]]
    \item Parallel describes the relationship among CSP\# atomic processes involved in parallel gateway. If Parallel (\textit{P1}) = [\textit{P2}[]], then the process must wait for atomic processes \textit{P1} and \textit{P2}[] finishing their execution to continue.
\end{itemize}

\begin{figure}[htbp]
    \centerline{\includegraphics[width=1\linewidth]{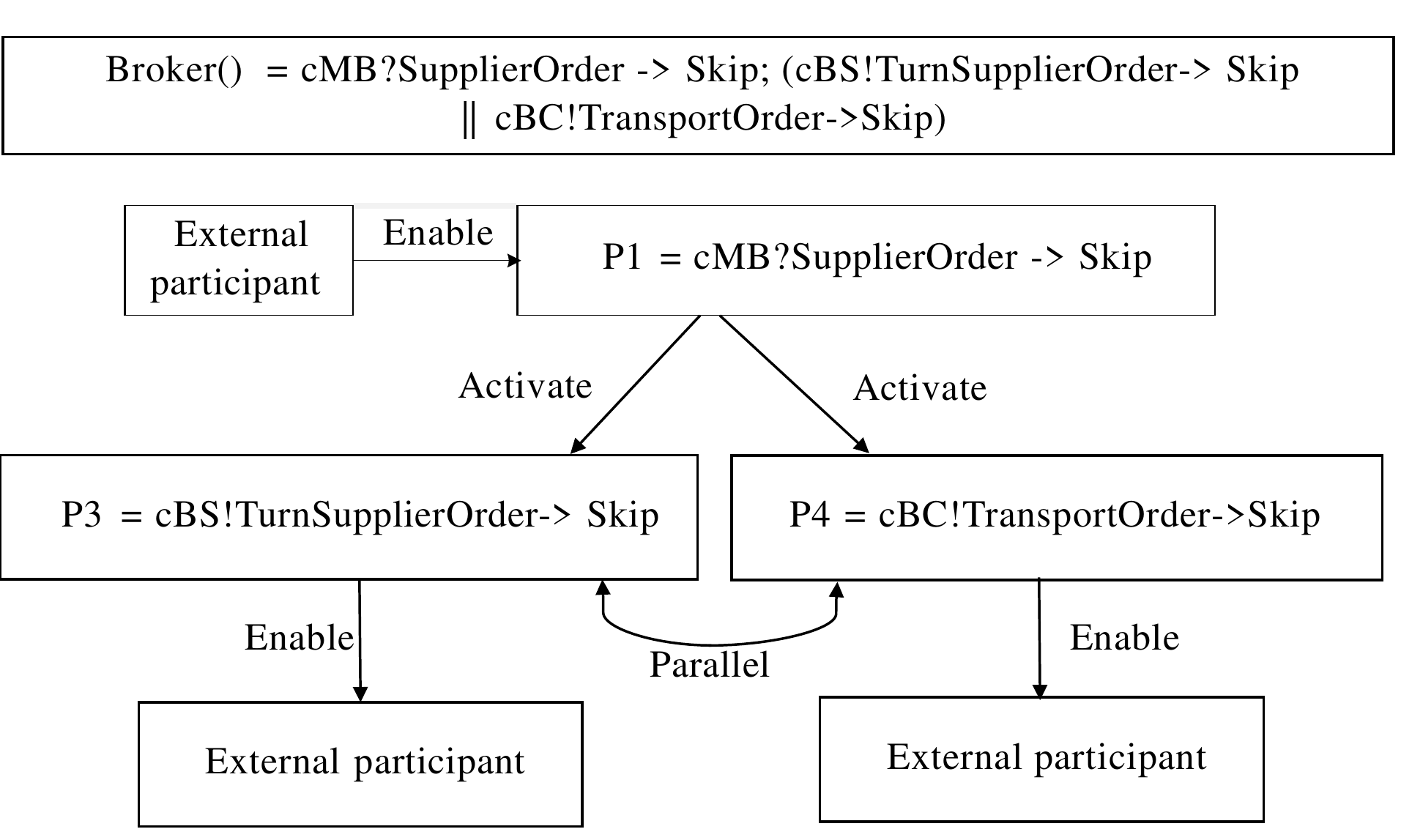}}
    \caption{New relationships of \textit{Broker} participant.}
    \label{fig:newfig4}
\end{figure}

As mentioned above, the new relationships only concern about atomic processes. The Reduction Algorithm helps us to get the new relationships by removing the state transitions involving the non-leaf nodes in the syntax trees. Now, we use Activate, Inactivate, Parallel and Enable to describe the interaction within and among participants. Fig.~\ref{fig:newfig4} shows the new relationships of \textit{Broker} participant. It can be observed that, compared with Fig.~\ref{fig:figure4}, the state transition is more simple, and composite processes like \textit{P2} have been removed.  

A part of the Reduction Algorithm is given here as an example. It starts with the CSP\# processes and the old relationships, ending with the new relationships. 

\begin{figure}[htbp]
    \centerline{\includegraphics[width=1\linewidth]{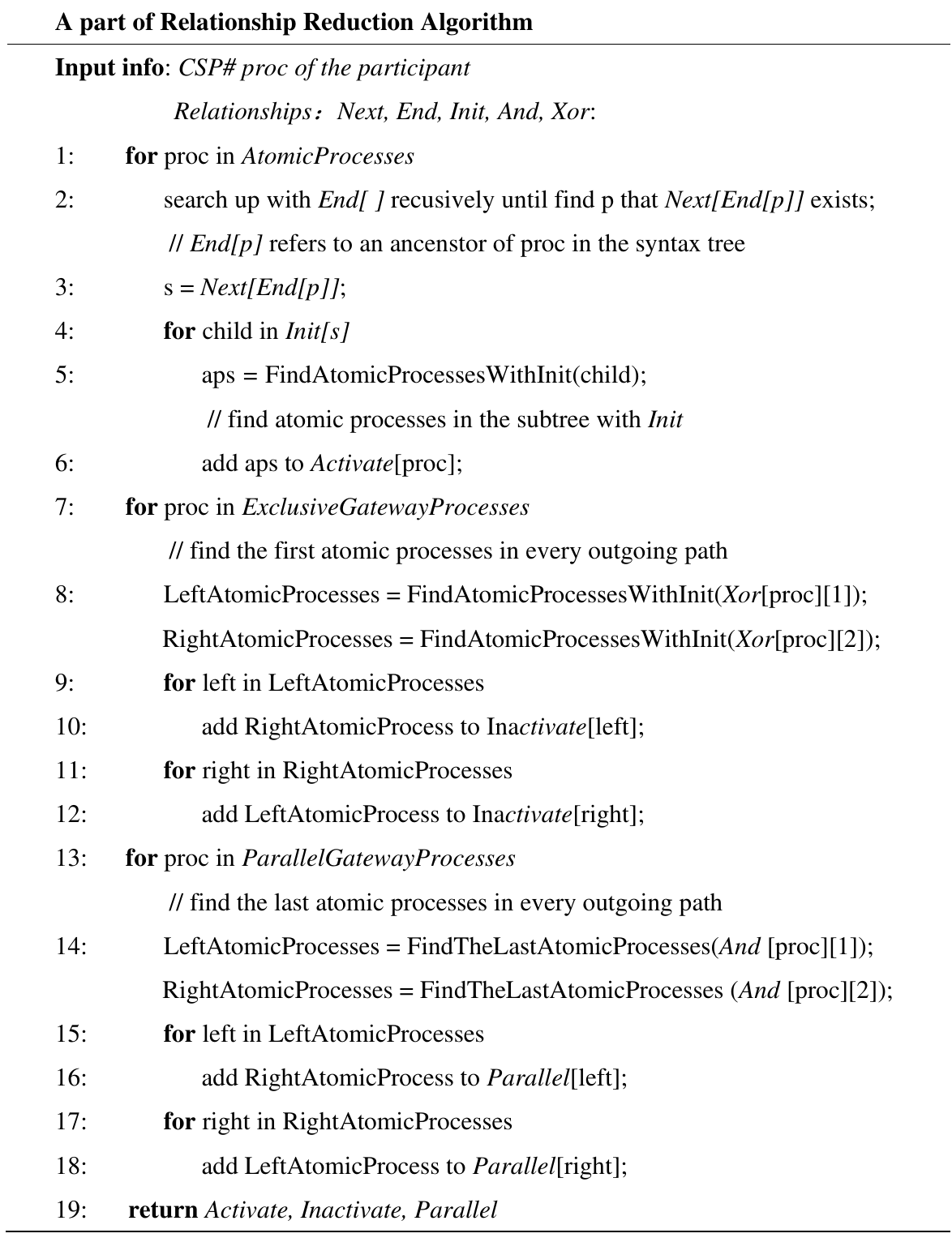}}
    \label{fig:reduction}
\end{figure}

\paragraph{Smart Contract Generation} 

Smart contract refers to the program code running in the blockchain environment. It controls the execution of the collaboration model in the order specified in advance. Specifically, it drives the change of process state by receiving external requests, judging the legitimacy of the request (e.g., whether the task can be executed), and sending requests to other participants. Here, we introduce the automatic generation method for Solidity smart contract that is widely used Ethereum environment.

The core of smart contract generation is to respond to the corresponding requests according to the states of CSP\# process and the relationships between CSP\# processes. These requests trigger changes in the states of CSP\# process. 

The states of CSP\# process are defined as follows. \textit{Disabled} indicates that the process is silent and execution is not allowed. \textit{Waiting} indicates that the process is enabled and waiting to be executed. \textit{Executing} indicates that the process is executing. \textit{Done} indicates that the execution of the process is completed and the process exits the executing state. 




In this paper, the smart contracts generated are composed of many functions used to handle external request.The external request is responsible for handling message interaction and event trigger. It checks the state of the CSP\# process and trigger the state change of other CSP\# processes according to the association relationships
At the same time, it forwards the message to the receiver (listening), and activates the receiving condition of the receiver. The external request algorithm is shown below.

\begin{figure}[htbp]
    \centerline{\includegraphics[width=1\linewidth]{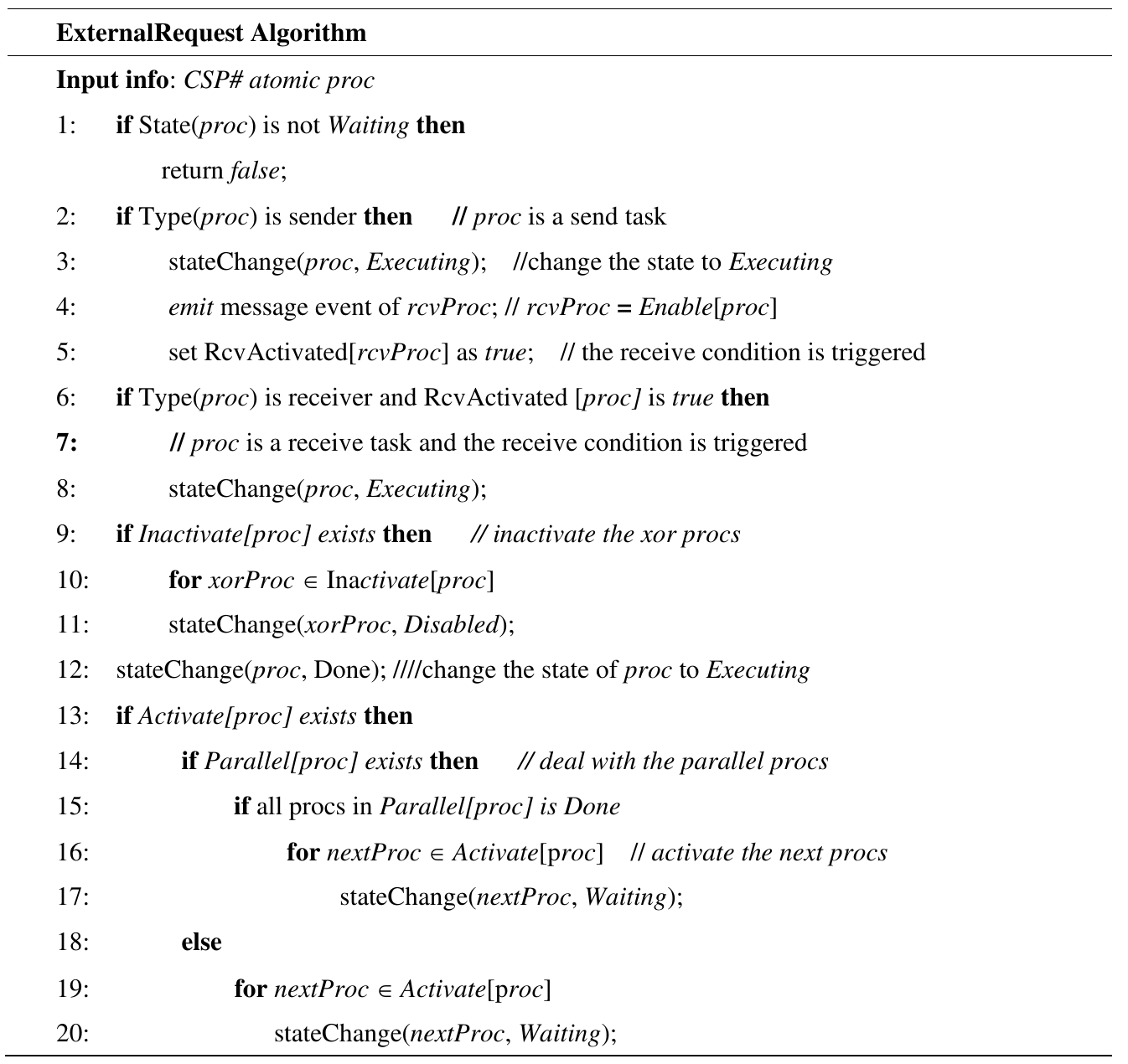}}
    \label{fig:figure7}
\end{figure} 

\section{Experiments}
This section tests a case set including 5 application cases to illustrate the effectiveness of the framework. These cases are from the existing literature \cite{ref4}, BPMN sample library and BPMAI \cite{ref17}, covering different application scenarios, namely supply chain (SC), booking travel (BT), online education (OE), paper review (PR) and pastry cook (PC). 
Here, we test each case (1) with smart contracts generated by our method(labelled as A). Besides, we  conduct a comparative experiment (2) with  contracts generated by the method of \cite{ref4}(labelled as B) and manually  contracts written in the way inspired by \cite{ref10}(labelled as C). Note that B is a classic method of smart contract automatic generation in the field of IOPC. This mainly considers the following reasons: Firstly, we want to observe whether the proposed framework meets our expected objectives (for (1)); Secondly, by quantitatively comparing the differences among the methods, we expect for a feasible direction for subsequent optimization (for (2)). Moreover, we provide a detailed analysis from the perspective of methodology in Section 6.

Table 2 shows the statistics of the experiments. Note that we have tested all the execution paths (branches) of each model and report the transaction data with the longest execution path in the order of message interaction.

\begin{table*}[]
    \centering
    \caption{Statistics of experiments}
    \begin{tabular}{|c|c|c|c|c|c|}
\hline
\multirow{2}{*}{\textbf{Case}}  & \textbf{Verification}   & \multicolumn{4}{|c|}{\textbf{Contract Execution}}    \\ \cline{2-6} 
                              & \textbf{Soundness(Y/N)} & \textbf{Item} & \textbf{Our Method(A)} & \textbf{B} & \textbf{C} \\ \hline
\begin{tabular}[c]{@{}c@{}}Supply Chain (SC)\\ (Round 1)\end{tabular}       & N     & ---   & ---   & ---   & ---        \\ \hline
\multirow{4}{*}{\begin{tabular}[c]{@{}c@{}}Supply Chain (SC)\\ (Round 2)\end{tabular}} & \multirow{4}{*}{Y}     
             & Total Gas Cost       & 1472682    & 1374716  & 1091035           \\ \cline{3-6} 
            &                         & Compared with C          & 135\%    & 126\%  & 100\%              \\\cline{3-6} 
            &                         & Initialization Gas Cost  & 1099479	&1016856 &767555           \\ \cline{3-6} 
            &                         & Initialization Gas Ratio & 74.66\%    & 73.97\%  & 70.35\%  \\ \cline{3-6}  \hline

\begin{tabular}[c]{@{}c@{}}Booking Travel (BT)\\ (Round 1)\end{tabular}       & N     & ---   & ---   & ---   & ---        \\ \hline
\multirow{4}{*}{\begin{tabular}[c]{@{}c@{}}Booking Travel (BT)\\ (Round 2)\end{tabular}} & \multirow{4}{*}{Y}     
             & Total Gas Cost       & 673808  & 653868    & 505198           \\ \cline{3-6} 
            &                         & Compared with C          & 133\%  & 129\%    & 100\%              \\\cline{3-6} 
            &                         & Initialization Gas Cost  & 566762 &548688 & 401313            \\ \cline{3-6} 
            &                         & Initialization Gas Ratio & 84.11\%    & 83.91\%  & 79.44\%  \\ \cline{3-6}  \hline  

\multirow{4}{*}{\begin{tabular}[c]{@{}c@{}}Online Education (OE)\end{tabular}} & \multirow{4}{*}{Y}     
             & Total Gas Cost       & 612334    & 567097  & 505198           \\ \cline{3-6} 
            &                         & Compared with C          &127\%    & 118\%  & 100\%              \\\cline{3-6} 
            &                         & Initialization Gas Cost  & 474703	& 434975	& 346800           \\ \cline{3-6} 
            &                         & Initialization Gas Ratio & 77.52\%    & 76.70\%  & 72.21\% \\ \cline{3-6}  \hline  
            
\multirow{4}{*}{\begin{tabular}[c]{@{}c@{}}Paper Review (PR)\end{tabular}} & \multirow{4}{*}{Y}     
             & Total Gas Cost       & 580467    & 600144  & 450903           \\ \cline{3-6} 
            &                         & Compared with C          & 128\%    & 133\%  & 100\%               \\\cline{3-6} 
            &                         & Initialization Gas Cost  & 475029	&494899	&347062            \\ \cline{3-6} 
            &                         & Initializations Gas Ratio & 81.84\%   &	82.46\%   &	76.97\%  \\ \cline{3-6}  \hline  
            
\multirow{4}{*}{\begin{tabular}[c]{@{}c@{}}Pastry Cook (PC)\end{tabular}} & \multirow{4}{*}{Y}     
             & Total Gas Cost      & 837937    & 741897  & 689443          \\ \cline{3-6} 
            &                         & Compared with C           & 121\%    & 107\%  & 100\%               \\\cline{3-6} 
            &                         & Initialization Gas Cost & 628723	&543578	&460708            \\ \cline{3-6} 
            &                         & Initialization Gas Ratio & 75.03\%   & 73.27\%    &	66.82\%  \\ \cline{3-6}  \hline  

    \end{tabular}
    \label{tab:table2}
    \end{table*}

One of advantages of our framework is that it provides model verification by introducing a formal model. Considering the Verification, the framework automatically performs model checking according to the selected properties. Once the verification fails, the framework will give corresponding counter examples to help users correct the problematic model. In these 5 cases, SC and BT are detected not to meet the soundness requirements, where Round 1 and
Round 2 in the table represent their first round verification
(failed) and modified verification (passed) respectively. If the verification fails, the model designers need to modify the model according to the prompt of counter example.

Only when the verification is passed, the framework generates the smart contract based on the formal model. After manual inspection one by one, all smart contracts comply with the original model specifications and can be executed correctly. The APPENDIX shows a segment of the smart contract
code of the supply chain scenario. 
However, in the comparative experiments, we find that it is not easy for experimenters to find abnormalities (because both methods don't include verification phase), and most of them still convert the problematic model directly into a smart contract. This reminds us that a formal model is necessary and meaningful as the input of smart contract generation. In this way, the verified formal model can not only provide clear execution semantics for the graphical model and eliminate ambiguity for the accurate generation of smart contracts, but also identify unqualified models in advance through model checking to avoid undesired contract implementation.

\parskip=0pt
It can be observed that in terms of Gas Cost, smart contracts generated by automation method cost more gas than manual written contracts. Both A (our method) and B bring more gas consumption than C, because smart contracts in A and B need to deal with more complex state transition conditions, leading to a lot of data recording costs and data update costs.

From the perspective of contract auto-generation, our method costs slightly more gas than B. However, our method could detect flaws of the graphic models in advance, which helps to avoid contract generation in the wrong base. 
Specifically, our method includes verification phase, and take as input the formal models that have passed the verification, rather than the manually designed graphic models that have a potential for flaws. In general, though our method costs 4\%-9\% more than B, we could save a lot of work of contract test, contract deploy and contract revision. 

Furthermore, when we compare the gas cost details of the contracts, we find that in all groups of our experiments, contract initialization, as the first transaction, is the most expensive one, always taking up more than 70\%. Taking the SC in the experiment A as an example, the gas cost of its first transaction is 1099479, which is much higher than the cost of subsequent transactions (the gas used is among 34019 and 56952). From this point of view, the main cost of the contract execution is contract initialization, while the gas cost caused by message interaction is relatively small. This reminds us that we should pay more attention to contract initialization in smart contract generation and optimization, so as to further reduce the cost.

\section{Related Work}
This section reviews the current work on smart contract generation and formalization of BPMN collaboration model. 
\subsection{Researches on Translation of BPMN Model to Smart Contract Generation}
The existing work mainly focuses on two different technical methods, namely the direct translation of BPMN models into smart contracts and the translation of BPMN models into smart contracts through intermediate formalisms. 

Due to the difference of input model types and framework configurations, it is difficult to provide a quantitative experimental analysis for different methods from time or cost. Here we make a comparative analysis from the aspects of method, formalism, input model and intention. We add reference \cite{ref8} because it is the latest representative on smart contract generation and is most relevant to ours in terms of intention.
Considering direct translation, Weber et al. \cite{ref4} for the first time propose the implementation and monitoring solution of IOPC based on blockchain to deal with the problem of mutual distrust among participants. To this end, it provides a translation algorithm and tool implemented in \cite{ref18} to map BPMN choreography elements to the corresponding smart contract. López-Pintado et al. \cite{ref5} combine the advantages of BPMS and blockchain platform to design a business process engine Caterpillar that can be executed on Ethereum. Different from other methods, it is a complete block-chain collaboration platform. Besides the smart contract of the collaboration model, other BPMS components such as work item are also embedded in the blockchain environment. 

Furthermore, the work of Ladleif et al. \cite{ref6} is an extension of \cite{ref4}. It points out the shortcomings of BPMN choreography model in terms of ownership and local observability according to the technical characteristics of blockchain (such as shared data and smart contract). Then the authors provide an extension and refinement of BPMN 2.0 choreography, and propose a proof of concept framework to fill the gap between modeling and implementation. From a model-driven perspective, Corradini et al. \cite{ref7} try to provide a bridge between the graphical model description and the low-level code executed on the blockchain. To this aim, the authors propose an implementation framework based on blockchain, as an infrastructure to support model management and Solidity smart contract generation.

In fact, compared with graphical models, formal models have advantages in accurately describing model behavior specifications. More importantly, it can use model checking to identify problematic models. These features help to avoid the generation of undesired smart contracts and enhance the confidence of blockchain-based IOPC (distributed systems) in software quality. Thus, some studies try to explore the indirect translation method through formal models to help optimize and improve the implementation of IOPC.
With respect to indirect translation, Zupan et al. \cite{ref8} point that smart contract is very error-prone, and it is difficult to repair the contract after implementation, and the formal model helps in advance discovery of threats that cause insecurity. The authors introduce Petri net into the smart contract generation to avoid unexpected problems. 

García-Bañuelos \cite{ref9} introduce Petri net as the intermediate carrier of translation. This work focuses on how to reduce the gas cost of transaction as much as possible while generating smart contract. It proposes an optimization method for contract initialization cost, execution cost and component throughput to reduce the impact of data volume and data update frequency on cost. It maps the BPMN model to a Petri net, and then simplifies the model by eliminating invisible transitions and redundant places. 

It is worth mentioning that Nakamura et al. \cite{ref10} introduce statechart into the automatic generation of smart contracts. It transforms a BPMN process with swimlanes into multiple statecharts. In this way, the statecharts can be simplified conveniently, thereby reducing the need for data exchange with the blockchain in collaboration. Then the statecharts as input are translated into the smart contracts.  

The work \cite{ref9} and \cite{ref10} are different from ours in terms of input model type and intention. Their focus is to reduce the cost of contract generation and execution by using the reduction technology of formal models, rather than focusing on the possible impact of problematic models. Furthermore, our work uses BPMN collaboration model to show the boundaries and roles of different participants. Meanwhile, in the formal method, we give a structured CSP\# formalization that considers message communication and process interaction, which simplifies the formal model and contract generation.

\subsection{Researches on CSP Formalization and Verification of BPMN Collaboration Model}
The formalization of BPMN model, and on a wider scale the formal study of IOPC, is a hot research field. Many works have explored from the perspective of formal language, e.g., PN/WF-net, Process algebra (Pi calculus), LTS, and First-Order Logic (FOL) and so on. As described in Section 2.2, CSP\# has advantages in process communication and structured representation, which help to reduce the complexity of smart contract generation. Here we focus on the existing CSP formalization and verification of BPMN collaboration related to our work.

As far as BPMN is concerned, both BPMN choreography and collaboration models could be used to model IOPC. The former is widely regarded as purely descriptive from a global point of view \cite{ref6}, while the latter has more advantages in intuitively representing the boundaries and business responsibilities of participants [3, p. 317]. More importantly, the BPMN collaboration has been widely used in the development of supporting software systems and serves as a starting point for model-driven development of distributed systems \cite{ref1}. Thus, we here choose the collaboration model to describe the participants and their interactions.

Wong et al. \cite{ref19} believe that a formal semantics for BPMN can ensure accurate specification and help designers to implement business processes correctly. To this end, the authors introduce Z syntax to describe BPMN structures. In particular, it employs the classic CSP and FOL to provide formalization for a subset of BPMN1.1. This formalization contains a lot of FOL and judgments (a lot of logic codes). This makes the formal models too complex and increases the complexity of smart contract generation.

Then, Corradini et al. \cite{ref20} focuses on specific e-Government digital service requirements, using CSP to represent collaboration model of BPMN 1.1. In this work, a task is considered to be an element that may directly associate multiple input and output sequence flows. Its corresponding formalization maps multiple parallel flow events (sequence flows) into a CSP process. This way may make the generated formalization too bloated. In our method, tasks, including sending and receiving types, focus on the task itself and ignore the internal flow that does not participate in the interaction, which helps to reduce the complexity of formalization. 

Furthermore, the work of Capel et al. \cite{ref21} can be regarded as an extension of \cite{ref19}. It uses Z syntax and CSP + Time to provide the formalization of Timed BPMN 2.0. Each model element is attached with time constraints or rules, which is mapped to the corresponding CSP process. This work focuses on specific time constraints and cannot be directly applied to smart contract generation. 

\section{Conclusion}
This paper proposes a prototype framework to automatically generate smart contracts for IOPC. It provides an end-to-end solution from modeling, verification, translation to implementation. As one of the cores, a CSP\# formalization for BPMN collaboration model bridges the graphical model with the smart contract. Another novelty is the translation algorithm of smart contract. It takes the verified formal model as input instead of graphical model, and translates it into Solidity smart contract based on syntax tree. The proposed framework identifies the abnormal model in advance through model checking, preventing undesired contract generation and implementation. In addition, the required formalism, verification and translation techniques are transparent to users without imposing additional burdens. In this regard, this reduces the complexity of studying blockchain-based distributed systems and model-driven software development. 

Some observations on gas cost are obtained from the experiment. Contract initialization is the most expensive, and its gas cost accounts for a high proportion of the total execution cost. This means that contract initialization should be an important optimization point when considering gas cost. In fact, as a hot topic, the combination of blockchain and IOPC is a systematic work involving many aspects, such as privacy, data storage, security and so on. This paper mainly discusses the relationship between IOPC described by BPMN collaboration models and smart contracts, and adopts the formal models to avoid the undesired contract implementations caused by low-quality models. 

In the future, reducing the cost of smart contract generation and execution is one of the directions we are interested in. In addition, we plan to optimize the framework architecture, providing more interfaces for external integration. And we also seek support for more complex elements, such as complex gateway and subprocesses.

\section*{Acknowledgment}

This work was supported by the National Key Research and Development Program of China under Grant No.2020YFB1707603; the NSFC-Guangdong Joint Fund Project under Grant No.U20A6003; the National Natural Science Foundation of China(NSFC) under Grant No.61972427; the Research Foundation of Science and Technology Plan Project in Guangdong Province under Grant No.2020A0505100030; the Research Foundation of Education Bureau of Hunan Province under Grant No.22B0669.


\bibliographystyle{IEEEtran}

\section*{Appendix}
A. This shows a part code of smart contract in the supply chain scenario
\begin{figure*}
    \centering
    \includegraphics[width=0.8\linewidth]{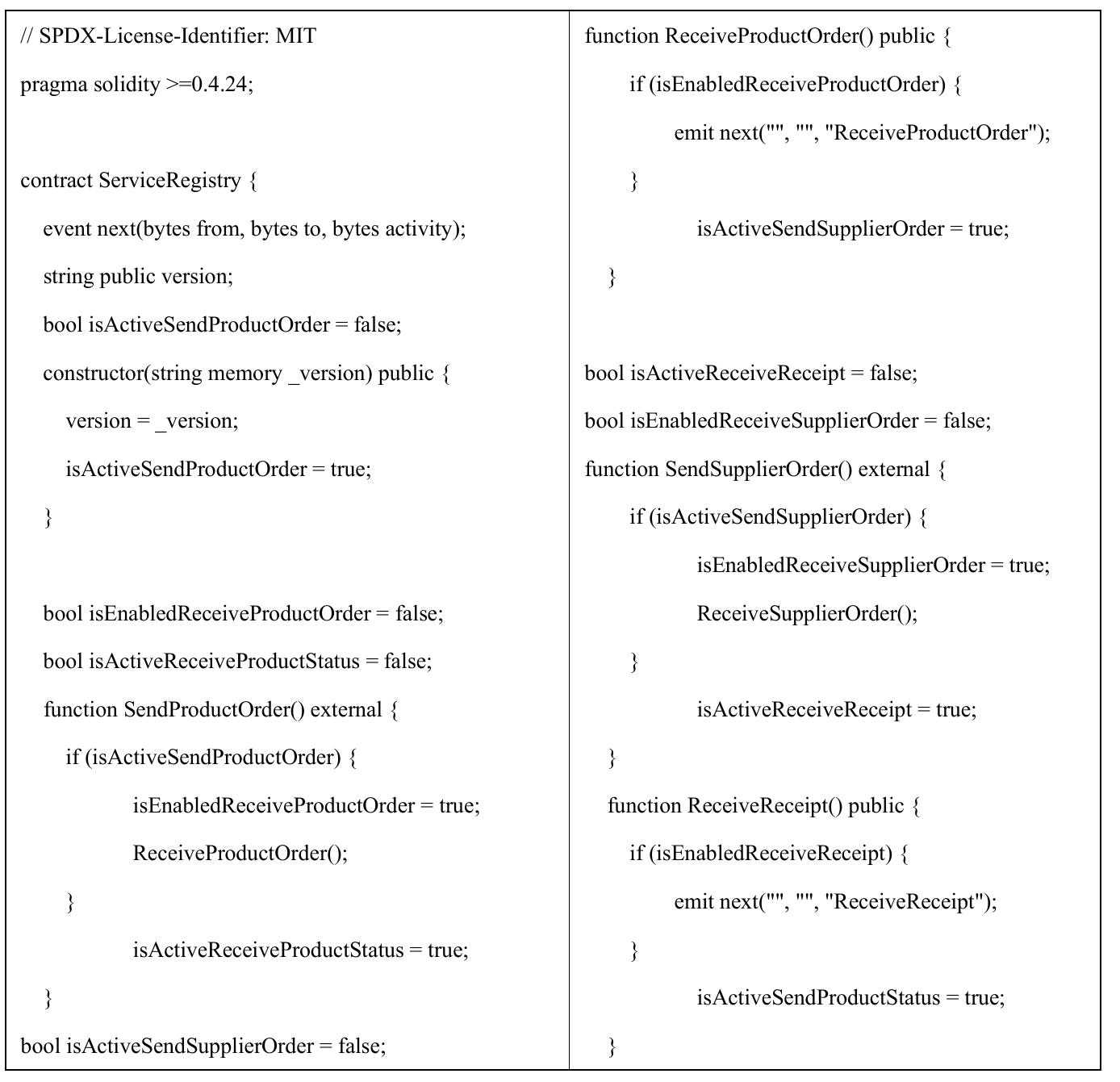}
    \label{fig:figure8}
  \end{figure*}
  
\end{document}